\documentclass[prd,aps,floats,preprintnumbers,preprint]{revtex4}

\usepackage{graphicx}
 
\newcommand{\be}{\begin{equation}}
\newcommand{\ee}{\end{equation}}
\newcommand{\bea}{\begin{eqnarray}}
\newcommand{\eea}{\end{eqnarray}}

\begin{document}

%\preprint{YITP-09-109}
%\preprint{KUNS-KUNS 2248}
%\preprint{RESCUE-28-09}

\setlength{\unitlength}{1mm}

\title{Corrections to the apparent value of the cosmological constant due to local inhomogeneities}

\author{Antonio Enea Romano$^{1,2,5}$}
\email{aer@phys.ntu.edu.tw}
\author{Pisin Chen$^{1,2,3,4}$}
\email{pisinchen@phys.ntu.edu.tw}
\affiliation{
${}^1$Department of Physics, National Taiwan University, Taipei 10617, Taiwan, R.O.C.\\
${}^2$Leung Center for Cosmology and Particle Astrophysics, National Taiwan University, Taipei 10617, Taiwan, R.O.C.\\
${}^3$Graduate Institute of Astrophysics, National Taiwan University, Taipei 10617, Taiwan, R.O.C.\\
${}^4$Kavli Institute for Particle Astrophysics and Cosmology, SLAC National Accelerator Laboratory, Menlo Park, CA 94025, U.S.A. \\
$^5$Instituto de Fisica, Universidad de Antioquia, A.A.1226, Medellin, Colombia\\
}

%\author{Antonio Enea Romano, Pisin Chen}
%\affiliation{
%$^1$Leung Center for Cosmology and Particle Astrophysics, National Taiwan University, Taipei %10617, Taiwan, R.O.C.\\
%$^2$Yukawa Institute for Theoretical Physics, Kyoto University,
%Kyoto 606-8502, Japan
%\\
%}

%\affiliation{
%Research Center for the Early Universe (RESCEU),
%Graduate School of Science, The University of Tokyo, Tokyo 113-0033, Japan}

\begin{abstract}
Supernovae observations strongly support the presence of a cosmological constant, but its value, which we will call apparent, is normally determined assuming that the Universe can be accurately described by a homogeneous model. Even in the presence of a cosmological constant  we cannot exclude nevertheless the presence of a small local inhomogeneity which could affect the apparent value of the cosmological constant. Neglecting the presence of the inhomogeneity can in fact introduce a systematic misinterpretation of cosmological data, leading to the distinction between an apparent and true value of the cosmological constant.
We establish the theoretical framework to calculate the corrections to the apparent value of the cosmological constant by modeling the local inhomogeneity with a $\Lambda LTB$ solution.
Our assumption to be at the center of a spherically symmetric inhomogeneous matter distribution correspond to effectively calculate the monopole contribution of the large scale inhomogeneities surrounding us, which we expect to be the dominant one, because of other observations supporting a high level of isotropy of the Universe around us. 

By performing a local Taylor expansion we analyze the number of independent degrees of freedom which determine the local shape of the inhomogeneity, and consider the issue of central smoothness, showing how the same correction can correspond to different inhomogeneity profiles.
Contrary to previous attempts to fit data using large void models our approach is quite general.  The correction to the apparent value of the cosmological constant is in fact present for local inhomogeneities of any size, and should always be taken appropriately into account both theoretically and observationally.  
\end{abstract}

\maketitle

\section{Introduction}
High redshift luminosity distance measurements \cite{Perlmutter:1999np,Riess:1998cb,Tonry:2003zg,Knop:2003iy,Barris:2003dq,Riess:2004nr} and
the WMAP measurements \cite{WMAP2003,Spergel:2006hy} of cosmic
microwave background (CMB) interpreted in the context of standard 
FLRW cosmological models have strongly disfavored a matter dominated universe,
 and strongly supported a dominant dark energy component, giving rise
to a positive cosmological acceleration.

As an alternative to dark energy, it has been 
proposed \cite{Nambu:2005zn,Kai:2006ws}
 that we may be at the center of an inhomogeneous isotropic universe without cosmological constant described by a Lemaitre-Tolman-Bondi (LTB)  solution of Einstein's field 
equations, where spatial averaging over one expanding and one contracting 
region is producing a positive averaged acceleration $a_D$, but it has been shown how spatial averaging can give rise to averaged quantities which are not observable \cite{Romano:2006yc}.
Another more general approach to map luminosity distance as a function of
 redshift $D_L(z)$ to LTB models has been recently
 proposed \cite{Chung:2006xh,Yoo:2008su},
 showing that an inversion method can be applied successfully to 
reproduce the observed $D_L(z)$.   
Interesting analysis of observational data in inhomogeneous models without dark energy and of other theoretically related problems is given for example in \cite{Alexander:2007xx,Alnes:2005rw,GarciaBellido:2008nz,GarciaBellido:2008gd,GarciaBellido:2008yq,February:2009pv,Uzan:2008qp,Quartin:2009xr,Quercellini:2009ni,Clarkson:2007bc,Ishibashi:2005sj,Clifton:2008hv,Celerier:2009sv,Romano:2007zz,Romano:2009qx,Romano:2009ej,Romano:2009mr,Mustapha:1998jb}

Here in this paper we will adopt a different approach. We will consider a Universe with a cosmological constant and some local large scale inhomogeneity modeled by a $\Lambda LTB$ solution \cite{Romano:2010nc}. For simplicity we will also assume that we are located at its center. In this regard this can be considered a first attempt to model local large scale inhomogeneities in the presence of the cosmological constant or, more in general, dark energy.
Given the spherical symmetry of the LTB solution and the assumption to be located at the center our calculation can be interpreted as the monopole contribution of the large inhomogeneities which surround us. Since we know from other observations such as CMB radiation that the Universe appears to be highly isotropic, we can safely assume that the monopole contribution we calculate should also be the dominant one, making our results even more relevant.
After calculating the null radial geodesics for a central observer we then compute the luminosity distance and compare it to that of $\Lambda CDM$ model, finding the relation between
the two different cosmological constants appearing in the two models, where we call apparent the one in the $\Lambda CDM$ and true the one in $\Lambda LTB$.
Our calculations show that the corrections to $\Omega^{app}_\Lambda$, which is the value of the cosmological constant obtained from analyzing supernovae data assuming homogeneity, can be important and should be taken into account.

%We need three functions to define a LTB solution, but because of the invariance under general coordinate %transformations, only two of them are really independent. This implies that two observables are in principle %sufficient to solve the inversion problem of mapping observations to a specific LTB model, for example the luminosity %distance $D_L(z)$ and the redshift spherical shell mass $m(z)n(z)=mn(z)$.

\section{LTB solution with a cosmological constant\label{ltb}}
The LTB solution can be written
as \cite{Lemaitre:1933qe,Tolman:1934za,Bondi:1947av} as
\begin{eqnarray}
\label{LTBmetric} %
ds^2 = -dt^2  + \frac{\left(R,_{r}\right)^2 dr^2}{1 + 2\,E(r)}+R^2
d\Omega^2 \, ,
\end{eqnarray}
where $R$ is a function of the time coordinate $t$ and the radial
coordinate $r$, $E(r)$ is an arbitrary function of $r$, and
$R_{,r}=\partial_rR(t,r)$.
The Einstein equations with dust and a cosmological constant give
\begin{eqnarray}
\label{eq2} \left({\frac{\dot{R}}{R}}\right)^2&=&\frac{2
E(r)}{R^2}+\frac{2M(r)}{R^3}+\frac{\Lambda}{3} \, , \\
\label{eq3} \rho(t,r)&=&\frac{2M,_{r}}{R^2 R,_{r}} \, ,
\end{eqnarray}
with $M(r)$ being an arbitrary function of $r$, $\dot
R=\partial_tR(t,r)$ and $c=8\pi G=1$ is assumed throughout the
paper. Since Eq. (\ref{eq2}) contains partial derivatives respect
to time only, its general solution can be obtained from the FLRW
equivalent solution by making every constant in the latter one an
arbitrary function of $r$.

The general analytical solution for a FLRW model with dust and cosmological
constant was obtained by Edwards \cite{Dilwyn} in terms of elliptic functions.
By an appropriate choice of variables and coordinates, we may extend it
to the LTB case thanks to the spherical symmetry of both LTB and FLRW models,
and to the fact that dust follows geodesics without being affected by
adjacent regions.
An anaytical solution can be found by introducing
a new coordinate $\eta=\eta(t,r)$ and a variable $a$ by \bea
\left(\frac{\partial\eta}{\partial
t}\right)_r=\frac{r}{R}\equiv\frac{1}{a}\,, \label{etadef} \eea
and new functions by \bea
\rho_0(r)\equiv\frac{6 M(r)}{r^3}\,,\quad
k(r)\equiv-\frac{2E(r)}{r^2}\,. \eea Then Eq. (\ref{eq2}) becomes
\be \left(\frac{\partial a}{\partial\eta}\right)^2 =-k(r)
a^2+\frac{\rho_0(r)}{3} a+\frac{\Lambda}{3} a^4\,, \ee where $a$ is now
regarded as a function of $\eta$ and $r$, $a=a(\eta,r)$. It should
be noted that the coordinate $\eta$, which is a generalization of
the conformal time in a homogeneous FLRW universe, has been only
implicitly defined by Eq.~(\ref{etadef}). The actual relation
between $t$ and $\eta$ can be obtained by integration once $a(\eta,r)$ is known:
\be
t(\eta,r)=\int_0^{\eta}{a(x,r)dx}+t_b(r) \,,
\ee
which can be computed analytically, and involve elliptic integrals of the third kind\cite{ellint}.

The function $t_B(r)$ plays the role of constant of integration, and is an arbitrary function of $r$, sometime called bang function, since by construction at time $t=t_b(r)$ we have $a(t_b(r),r)=0$, and correspond to the fact that the big bang initial singularity can happen at different times at different positions from the center in a LTB space. In the rest of this paper we will assume homogeneous bang, i.e. we will set
\be
t_b(r)=0.
\ee
Inspired by the construction of the solution for the FLRW case get:
\be
a(\eta,r)
=\frac{\rho_0(r)}{3\phi\left(\frac{\eta}{2 };g_2(r),g_3(r)\right)+k(r)}\,,
\ee
where $\phi(x;g_2,g_3)$ is the Weierstrass elliptic function
satisfying the differential equation 
\be
\left(\frac{d\phi}{dx}\right)^2=4\phi^3-g_2\phi-g_3\, ,\label{eqweir} 
\ee 
and
\bea
\alpha=\rho_0(r)\,,\quad
g_2=\frac{4}{3}k(r)^2 \,,\quad
g_3=\frac{4}{27} \left(2k(r)^3 -\Lambda\rho_0(r)^2\right)\,.
\eea
In this paper we will choose the so called FLRW gauge, i.e. the coordinate system in which $\rho_0(r)$ is constant. 

\section {Geodesic equations and luminosity distance}

We adopt the same method developed in \cite{Romano:2009xw} to
solve the null geodesic equation written in terms of the
coordinates $(\eta,r)$. Instead of integrating differential
equations numerically, we perform a local expansion of the
solution around $z=0$ corresponding to the point $(t_0,0)$, or
equivalently $(\eta_0,0)$, where $t_0=t(\eta_0,0)$. The change of
variables from $(t,r)$ to $(\eta,r)$ permits us to have r.h.s. of
all equations in a fully analytical form, in contrast to previous
considerations of this problem which require a numerical
calculation of $R(t,r)$ from the Einstein equation~(\ref{eq2}).
Thus, this formulation is particularly suitable for derivation of
analytical results.

The luminosity distance for a central observer in the LTB
space-time as a function of the redshift $z$ is expressed as \be
D_L(z)=(1+z)^2 R\left(t(z),r(z)\right) =(1+z)^2
r(z)a\left(\eta(z),r(z)\right) \,, \ee where
$\Bigl(t(z),r(z)\Bigr)$ or $\Bigl((\eta(z),r(z)\Bigr)$ is the
solution of the radial geodesic equation as a function of $z$. The
past-directed radial null geodesics is given by \bea \label{geo1}
\frac{dt}{dr}
%=f(T(r),r) \,; \quad f(t,r)
=-\frac{R_{,r}(t,r)}{\sqrt{1+2E(r)}} \,. \eea
%where $t=T(r)$ is the time coordinate along the null radial geodesic as a
%function of the coordinate $r$. From the implicit solution, we can
%rewrite the above as \bea T(r)=t(U(r),r)\,; \quad
%\frac{dT(r)}{dr}= \frac{\partial t}{\partial \eta}
%\frac{dU(r)}{dr}+\frac{\partial t}{\partial r}\,, \eea where
%$\eta=U(r)$ is the $\eta$-coordinate along the null radial
%geodesic as a function of $r$.
In terms of $z$, Eq. (\ref{geo1}) takes the form
 \cite{Celerier:1999hp}:
\begin{eqnarray}
{dr\over dz}&=&{\sqrt{1+2E(r(z))}\over {(1+z){\dot R}_{,r}[r(z),t(z)]}} \,,
\nonumber\\
{dt\over dz}&=&-{R_{,r}[r(z),t(z)]\over {(1+z){\dot
R}_{,r}[r(z),t(z)]}} \,. \label{eq:35}
\end{eqnarray}
%These equations are derived from the definition of redshift and
%by following the evolution of a short time interval along the null
%geodesic $T(r)$.
The inconvenience of using the $(t,r)$ coordinates is that there
is no exact analytical solution for $R(t,r)$. So the r.h.s. of
Eqs. (\ref{eq:35}) cannot be evaluated analytically, but we are
required to find a numerical solution for $R$ first
\cite{Hellaby:2009vz}, and then to integrate numerically the
differential equations, which is quite an inconvenient and
cumbersome procedure, and cannot be used to derive anaytical results.

It can be shown \cite{Romano:2009xw} that in the coordinates $(\eta,r)$ eqs. (\ref{eq:35}) take the form: 
\bea \label{geo3}
\frac{d \eta}{dz} &=&-\frac{\partial_r
t(\eta,r)+F(\eta,r)}{(1+z)\partial_{\eta}F(\eta,r)}
\equiv p(\eta,r) \,,\\
\label{geo4} \frac{dr}{dz}
&=&\frac{a(\eta,r)}{(1+z)\partial_{\eta}F(\eta,r)} \equiv
q(\eta,r) \,, \eea where \be F(\eta,r)\equiv \frac{\
R_{,r}}{\sqrt{1+2E(r)}}=
\frac{1}{\sqrt{1-k(r)r^2}}\left[\partial_r (a(\eta,r) r)
-a^{-1}\partial_{\eta} (a(\eta,r) r)\, \partial_r t(\eta,r)\right]
\,. \ee
%where $\eta=U(r(z))$ and $F(\eta,r)=f(t(\eta,r),r)$.
It is important to observe that the functions $p,q,F$ have explicit
analytical forms, making it particularly useful do derive anaytical results.

%which can be obtained from $a(\eta,r)$ and $t(\eta,r)$ as shown below.
%The derivation of the implicit  solution $a(\eta,r)$ is based on the use of
%the conformal time variable $\eta$, which by construction
%satisfies the relation,
%\be
%\frac{\partial\eta(t,r)}{\partial t}=a^{-1} \,.
%\ee
%This means
%In order to use the analytical solution we need to find analytical
%expressions for $F$ and $F_{,\eta}$.
% This can always be done by using
%\bea
%\frac{\partial}{\partial t}=a^{-1}{\frac{\partial}{\partial \eta}}\,,
%\quad
%\partial_r \eta=
%-a(\eta,r)^{-1}\partial_r t  \, .
%\eea
\section{Number of independent parameters and Taylor expansion accuracy}
In order to find the relation between the apparent and true value of the cosmological constant we need in to match the terms in the red-shift expansion :
\bea
D_i^{\Lambda CDM}=D^{{\Lambda LTB}}_i& ,
\eea
Before proceeding in deriving this relation we need to understand clearly how many independent parameters we can solve for at different order in the Taylor expansion for $D_L(z)$.
After defining the expansion of the function $k(r)$ in terms of the dimensionless function $K(r)$:
\be
k(r)=(a_0 H_0)^2 K(r)=K_0+K_1 r+K_2 r^2+ ..\\
\ee
we have
\bea
D_1^{\Lambda LTB}&=&\frac{1}{H^{\Lambda LTB}_0} \,, \\
D^{\Lambda LTB}_{i}&=&f_i(\Omega_{\Lambda},K_0,K_1,..,K_{i-1}),
\eea
which implies that if we want to match the coefficient $D_i$ up to order n, we will have a total of $n+2$ independent parameters to solve for :
\be
\{H^{\Lambda}_0,\Omega_{\Lambda},K_0,K_i,..,K_{i-1}\}.
\ee 
The matching conditions will imply a constraint over the $n+2$ independent parameters, but this will not be enough completely determine them, since two of them will always be free.
For a matter of computational convenience we will choose $K_0,K_1$ as free parameters and express all the other in terms of them.
For example from :
\be
D_2^{\Lambda CDM}=D^{{\Lambda LTB}}_2, \label{eqD2}
\ee
we can get
\be
\Omega^{app}_{\Lambda}(\Omega_{\Lambda},K_0,K_1),
\ee
from
\be
D_3^{\Lambda CDM}=D^{{\Lambda LTB}}_3,
\ee
we can get 
\be
K_2(\Omega_{\Lambda}^{app},K_0,K_1), 
\ee
and in general from
\be
D_i^{\Lambda CDM}=D^{{\Lambda LTB}}_i,
\ee
we can get 
\be
K_{i-1}(\Omega_{\Lambda}^{app},K_0,K_1). 
\ee
Since our purpose is to find the corrections to the apparent value of the cosmological constant, the second order term $D_2$ is enough. Higher order terms in the redshift expansion will provide $K_2,K_3,..K_{i-1}$ as functions of $\{\Omega_{\Lambda}^{app},K_0,K_1\}$, but will not change the analytical relation between $\Omega^{app}_{\Lambda}$ and $\Omega_{\Lambda}$ which can be derived from eq.(\ref{eqD2}).
For this reason we will only need the expansion up to second order for the luminosity distance.
The fact that we have more free parameters than constraints implies that the same correction to the apparent value of the cosmological constant can correspond to an infinite number of different inhomogeneity profiles.

The corrections we calculate are accurate within the limits of validity of the Taylor expansion $D^{\Lambda CDM}_{Taylor}$. It turns out that in the flat case we consider the error is quite large already at a redshift of about $0.2$ as shown in the figure.
This implies that the corrections should also be valid only within this low redshift range, since even if we are exactly matching the coefficients, the Taylor expansion of the $\Lambda CDM$ best fit formula itself is not very accurate.
This could be overcome by implementing other types of expansions or numerical methods, such as Pad\'{e} for example, with better convergence behavior, but we'll leave this to a future work.

\begin{figure}[t]
\includegraphics[height=55mm,width=80mm]{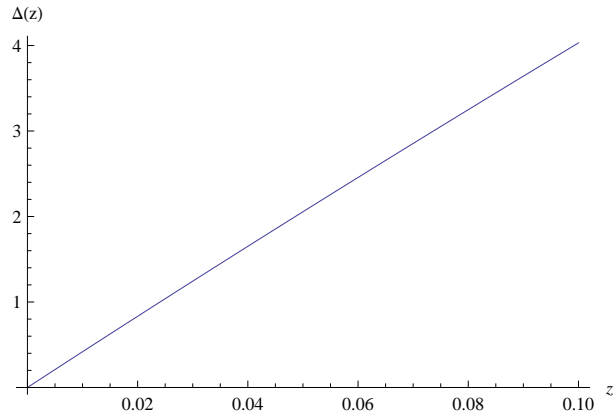}
\caption{The percentual error $\Delta=100\frac{ D^{\Lambda CDM}-D^{\Lambda CDM}_{Taylor} }{D^{\Lambda CDM}}$ for a third order expansion is plotted as a function of the redshift.
As it can be seen the error is already quite large at redshift 0.1. Higher order expansion does not improve the convergence.}
\label{krI}
\end{figure}

\section{Central behavior}
A function of the radial coordinate $f(r)$ is smooth at the center $r=0$ only if all its odd derivatives vanish there. 
This can be shown easily by looking at the partial derivatives of even order of this type for example:
\bea
\partial^{2n}_x\partial^{2n}_y\partial^{2n}_z f(\sqrt{x^2+y^2+z^2}) \,,
\eea
where $\{x,y,z\}$ are the cartesian coordinates related to $r$ by $r^2=x^2+y^2+z^2$.
Quantities of the type above diverge at the center if $\partial^{2m+1}_r f(r) \neq0$ for $2m+1<2n$.
If for example the first derivative $f'(0)$ is not zero, then the laplacian will diverge.
This implies that including linear terms expansions for $k(r)$ and $t_b(r)$ we are considering models which are not smooth at the center.
The general central smoothness conditions are:
\bea
k_{2m+1}&=&0, \\
t_b^{2m+1}&=&0 \,, \\
2m+1&<&i \,,
\eea
which must be satisfied for all the relevant odd powers coefficients of the central Taylor expansion.
In our case this implies that if we only want to consider centrally smooth inhomogeneities then we need to set to zero all the odd derivatives of $K(r)$
\be
K_{2m+1}=0
\ee
The consequence of this smoothness conditions is that the exact matching of the Taylor expansion is possible only up to order five when we have five constraints equations
\bea
D_i^{\Lambda CDM}=D^{{\Lambda LTB}}_i& \quad ,\quad 1\leq i\leq 5\,,
\eea
and five free parameters
\bea
{H_0^{\Lambda LTB},\Omega_{\Lambda},K_0,K_2,K_4}
\eea
implying there is a unique solution.
Going to higher order there will be more equations than free parameters making the inversion problem impossible.
This means that the effects of a different value of the cosmological constant cannot be mimicked by a smooth inhomogeneity, as far as the exact matching of the Taylor expansion is concerned.
From a data analysis point of view this limitation could be easily circumvented, since these considerations are based on matching the Taylor expansion of the best $\Lambda CDM$ fit, which is quite different from fitting the actual data.
Also it turns out that the Taylor expansion $D^{\Lambda CDM}_{Taylor}(z)$ is more accurate at second order than at any other order as shown in the figure, implying that exact matching beyond second order is practically irrelevant from a data fitting point of view. 
Under these considerations the inversion problem can be considered still effectively undetermined since by matching up to second order we have two equations and three parameters:
\bea
{H_0^{\Lambda LTB},\Omega_{\Lambda},K_0}
\eea
For completeness of the analysis we mention that after counting the number of independent parameters we can easily conclude that the inversion problem remain undetermined for the third order, and has a unique solution for the fourth and fifth order as shown above.

\section {Calculating the luminosity distance}
In order to obtain the redshift expansion of the luminosity distance we need to use the following: 
\bea
k(r)&=&(a_0 H_0)^2 K(r)=K_0+K_1 r+K_2 r^2+ ..\\
t(\eta,r)&=& b_0(\eta)+b_1(\eta)r+b_2(\eta)r^2 + ..
\eea
It should be noted that linear terms will in fact lead to central divergences of the laplacian in spherical coordinates, which correspond to a central spike of the energy distribution   \cite{Romano:2009ej,Romano:2009mr}, but an appropriate local averaging of the solution can easily heal this behavior, and we include them here because they give the leading order contribution.
Since we are interested in the effects due to the inhomogeneities we will neglect $k_0$ in the rest of the calculation because this corresponds to the homogeneous component of the curvature function $k(r)$.

Following the same approach given in \cite{Romano:2010nc} , we can find a local Taylor expansion in red-shift for the geodesics equations, and then calculate the luminosity distance:
\bea
D^{\Lambda LTB}_L(z)&=&(1+z)^2r(z)a^{\Lambda LTB}(\eta(z),r(z))=D^{\Lambda LTB}_1 z+D^{\Lambda LTB}_2 z^2+D^{\Lambda LTB}_3 z^3 + . .\\
D^{\Lambda LTB}_1&=&\frac{1}{H_0},\nonumber\\
D^{\Lambda LTB}_2&=&\frac{1}{{36 H_0 (\Omega_{\Lambda}^{true}-1)}}\bigg[54 B_1 (\Omega_{\Lambda}^{true}-1)^2+18 B'_1 (\Omega_{\Lambda}^{true}-1)-18 h_{0,r} (\Omega_{\Lambda}^{true})^2\nonumber \\
&&+30 h_{0,r} \Omega_{\Lambda}^{true}-12 h_{0,r}+6K_1 \Omega_{\Lambda}^{true}-10 K_1+27 (\Omega_{\Lambda}^{true})^2-18 \Omega_{\Lambda}^{true}-9\bigg],
%D_2&=&\frac{1}{{2 a_0^3 H_0^4 (2 q_0-1)^{5/2}}}\bigg[\sqrt{2 q_0-1} (-a_0^3 H_0^3 (1-2 q_0)^2 %(q_0-1)-2 a_0^2 H_0^3 (4
%   q_0^3-3 q_0+1) t^b_1-9 k_1 q_0)+\nonumber\\
%   &&+6 k_1 q_0 (q_0+1) X\bigg]\nonumber\\
%D_3&=&\frac{}{{4 a_0^6
%   H_0^7 (2 q_0-1)^{11/2}}}\bigg[-3 (2 q_0-1) q_0 X (4 a_0^2 H_0^3
%   k_1 (1-2 q_0)^2 q_0 (2 a_0 q_0+4 q_0 t^b_1+t^b_1)+\nonumber \\
%   &&-4 a_0^2 H_0^2
%   k_2 (4 q_0^3-3 q_0+1)+k_1^2 (50 q_0^2+7 q_0-7))+(2
%   q_0-1)^{3/2} (2 a_0^6 H_0^6 (1-2 q_0)^4 (q_0-1) q_0+\nonumber\\
%   &&+8 a_0^5 H_0^6 (1-2
%   q_0)^4 q_0^2 t^b_1+4 a_0^3 H_0^3 k_1 (1-2 q_0)^2 q_0 (5 q_0-1)+\nonumber\\
%   &&+2a_0^2 H_0^2 (1-2 q_0)^2 (H_0 k_1 (20 q_0^2+q_0-1) t^b_1-9
%   k_2 q_0)+2 H_0^5 (a_0-2 a_0 q_0)^4 (H_0 q_0 (4 q_0+1)
%   {(t^b_1)}^2+\nonumber\\
%   &&-2 (q_0+1) t^b_2)+9 k_1^2 q_0 (11 q_0-4))+18 k_1^2 \sqrt{2
%   q_0-1} (4 q_0+1) q_0^3 X^2\bigg]  \nonumber
\eea
where we have introduced the dimensionless quantities $K_0,K_1,B_1,B'_1,h_{0,r}$ according to
\bea
H_0&=&\left(\frac{\partial_t, a(t,r)}{a(t,r)}\right)^2\Biggr|_{t=t_0,r=0}=\left(\frac{\partial_{\eta} a(\eta,r)}{a(\eta,r)^2}\right)^2\Biggr|_{\eta=\eta_0,r=0}, \\
%K_0&=&k_0(a_0 H_0)^{-2}, \\
%K_1&=&k_1(a_0 H_0)^{-3}, \\
B_1(\eta)&=&b_1(\eta) a_0^{-1},  \\
B_1&=&b_1(\eta_0) a_0^{-1}, \\
B'_1&=&\frac{\partial B_1(\eta)}{\partial \eta}\Biggr|_{\eta=\eta_0} (a_0 H_0)^{-2}, \\
h_{0,r}&=&\frac{1}{a_0 H_0}\frac{\partial_r a(\eta,r)}{a(\eta,r)}\Biggr|_{\eta=\eta_0,r=0}, \\
t_0&=&t(\eta_0,0),
\eea
and used the Einstein equation at the center $(\eta=\eta_0,r=0)$
\bea
1&=&\Omega_k(0)+\Omega_M+\Omega_{\Lambda}=-K_0+\Omega_M+\Omega_{\Lambda},\\
\Omega_k(r)&=&-\frac{k(r)}{H_0^2 a_0^2},\\
\Omega_M&=&\frac{\rho_0}{3 H_0^2 a_0^3},\\
\Omega_{\Lambda}&=&\frac{\Lambda}{3 H_0^2}.
\eea
Because of our coordinate choice $\Omega_M$ is independent of $r$, and all the radial dependence goes into $\Omega_k(r)$. 
Note that apart from the central curvature term $K_0$, the inhomogeneity of the $LTB$ space is expressed in $h_{0,r}$, which encodes the radial dependence of the scale factor.
Details of these rather cumbersome calculations are provided in a separate companion paper,  but it should be emphasized that in order to put the formula for the luminosity distance in this form it is necessary to manipulate appropriately the elliptic functions and then re-express everything in terms of physically meaningful quantities such as $H_0$.

\section{Calculating $D_L(z)$ for $\Lambda CDM$ models.}
The metric of a $\Lambda CDM$ model is the FLRW metric, a special case of LTB solution, where :
\bea
\rho_0(r)&\propto& const,\\
k(r)&=&0, \\
t_b(r) &=&0, \\
a(t,r)&=&a(t).
\eea
We will calculate independently the expansion of the luminosity distance and the redshift spherical shell mass for the case of a flat $\Lambda CDM$, to clearly show the meaning of our notation, and in particular the distinction between $\Omega_{\Lambda}^{app}$ and $\Omega_{\Lambda}^{true}$.
We can also use these formulas to check the results derive before, since in absence of inhomogeneities they should coincide.

One of the Einstein equation can be expressed as:
\bea
H^{\Lambda CDM}(z)&=&H_0\sqrt{(1-\Omega_{\Lambda}^{app}){\left(\frac{a_0}{a}\right)}^3+\Omega_{\Lambda}^{app}}=H_0\sqrt{(1-\Omega_{\Lambda}^{app}){(1+z)}^3+\Omega_{\Lambda}^{app}}.
\eea
We can then calculate the luminosity distance using the following relation, which is only valid assuming flatness:
\bea
D^{\Lambda CDM}_L(z)=(1+z)\int^z_0{\frac{d z'}{H^{\Lambda CDM}(z')}}=D^{\Lambda CDM}_1 z+D^{\Lambda CDM}_2 z^2+D^{\Lambda CDM}_3 z^3+ . . .
\eea
From which we can get:
\bea
D^{\Lambda CDM}_1&=&\frac{1}{H_0}\,,\\
D^{\Lambda CDM}_2&=&\frac{3 \Omega_{\Lambda}^{app}+1}{4 H_0}\,.
\eea
We can check the consistency between these formulae and the ones derived in the case of LTB by setting:
\bea
K_1=B_1=B'_1=K_0=h_{0,r}=0\,,\\
\eea
which corresponds to the case in which $\Omega_{\Lambda}^{app}=\Omega_{\Lambda}^{true}$.

\section{Relation between apparent and true value of the cosmological constant}
So far we have calculated the first two terms of the redshift expansion of the luminosity distance for $\Lambda LTB$ and $\Lambda CDM$ model. Since we now that the latter provides a good fitting for supernovae observations, we can now look for the $\Lambda LTB$ models which give the same theoretical prediction. From the above relations we can  derive :
\bea
H_0^{\Lambda LTB}&=&H^{\Lambda CDM}_0\,, \\
\Omega_{\Lambda}^{app}&=&\frac{1}{{27
   (\Omega_{\Lambda}^{true}-1)}}\Bigg[54 B_1 (\Omega_{\Lambda}^{true})^2-108 B_1 \Omega_{\Lambda}^{true}+54 B_1+18 B'_1 \Omega_{\Lambda}^{true}-18 B'_1\nonumber \\
   &&-18 h_{0,r}
   (\Omega_{\Lambda}^{true})^2+30 h_{0,r} \Omega_{\Lambda}^{true}-12 h_{0,r}+6 K_1 \Omega_{\Lambda}^{true}-10 K_1\nonumber\\
   &&+27 \Omega_{\Lambda}^{true}(\Omega_{\Lambda}^{true}-1)\Bigg]\,, \\   
\Omega_{\Lambda}^{true}&=&-\frac{1}{{6 (6 B_1-2 h_{0,r}+3)}}\Bigg[\Bigg((36 B_1-6 B'_1-10 h_{0,r}-2 K_1+9 \Omega_{\Lambda}^{app}+9)^2 +\nonumber\\
&&-4 (6 B_1-2 h_{0,r}+3)(54
   B_1-18 B'_1-12 h_{0,r}-10 K_1+27 \Omega_{\Lambda}^{app})\Bigg)^{1/2}-36 B_1\nonumber\\
   &&+6 B'_1+10 h_{0,r}+2 K_1-9(\Omega_{\Lambda}^{app}-1)\Bigg]\,.
\eea
We can also expand the above exact relations assuming that all the inhomogeneities, can be treated perturbatively respect to the $\Lambda CDM $ , i.e. $\{{K_1,B_1,B'_1\}}\propto \epsilon $, where $\epsilon$ stands for a small deviation from $FLRW$ solution :
\bea
%\Omega_{\Lambda}^{app}&=& \Omega_{\Lambda}^{true}+\frac{2  \left(27 B_1 (\Omega_{\Lambda}^{true}-1)^2+9 B'_1 %(\Omega_{\Lambda}^{true}-1)-9 h_{0,r} (\Omega_{\Lambda}^{true})^2+15
%   h_{0,r} \Omega_{\Lambda}^{true}-6 h_{0,r}+3 K_1 \Omega_{\Lambda}^{true}-5 K_1\right)}{27 %(\Omega_{\Lambda}^{true}-1)}+O\left(\epsilon ^2\right)\\
\Omega_{\Lambda}^{true}&=&\Omega_{\Lambda}^{app}-\frac{2}{{27 (\Omega_{\Lambda}^{app}-1)}}(27 B_1 (\Omega_{\Lambda}^{app}-1)^2+9 B'_1 (\Omega_{\Lambda}^{app}-1)-9 h_{0,r} {(\Omega_{\Lambda}^{app})}^2+15
   h_{0,r} \Omega_{\Lambda}^{app}\nonumber \\
   &&-6 h_{0,r}+3 K_1 \Omega_{\Lambda}^{app}-5 K_1)+O(\epsilon ^2)\,.
\eea
As expected all these relations reduce to 
\bea
\Omega_{\Lambda}^{true}&=&\Omega_{\Lambda}^{app},
\eea
in the limit in which there is no inhomogeneity, i.e. when $K_1=B_1=B'_1=h_{0,r}=0$.

\section{Conclusions}

We have derived for the first time the correction due to local large scale inhomogeneities to the value of the apparent cosmological constant inferred from low redshift supernovae observations.
This analytical calculation shows how the presence of a local inhomogeneity can affect the estimation of the value of cosmological parameters, such as $\Omega_{\Lambda}$. This effects should be properly taken into account both theoretically and observationally.
By performing a local Taylor expansion we analyzed the number of independent degrees of freedom which determine the local shape of the inhomogeneity, and consider the issue of central smoothness, showing how the same correction can correspond to different inhomogeneity profiles.
We will address in a future work the estimation of the magnitude of this effect based on experimental bounds which can be set on the size and shape of a local inhomogeneity and the fitting of actual supernovae data. It is important to underline here that we do not need a large void as normally assumed in previous studies of $LTB$ models in a cosmological context. Even a small inhomogeneity could in fact be important. 

In the future it will also be interesting to extend the same analysis to other observables such as barionic acoustic oscillations (BAO) or the cosmic microwave background radiation (CMBR), and we will report about this in separate papers.
Another direction in which the present work could be extended is modeling the local inhomogeneity in a more general way, for example considering not spherically symmetric solutions.
From this point of view our calculation could be considered the  monopole contribution to the general effect due to a local large scale inhomogeneity of arbitrary shape. 
Given the high level of isotropy of the Universe shown by other observations such as the CMB radiation, we can expect the monopole contribution we calculated to be the dominant one.

While this should be considered only as the first step towards a full inclusion of the effects of large scale inhomogeneities in the interpretation of cosmological observations, it is important to emphasize that we have introduced a general definition of the concept of apparent and true value of cosmological parameters, and shown the general theoretical approach to calculate the corrections to the apparent values obtained under the standard assumption of homogeneity.

%While this should be considered only the fist step towards of full inclusion of the effects of %large scale inhomogeneities in the interpretation of cosmological observations, it is %important to emphasize that we have clearly introduced the notion of apparent and true value %of cosmological parameters, and of the cosmological constant in particular, and shown the %obtained under the standard assumption of homogeneity.

\begin{acknowledgments}
We thank A. Starobinsky, M. Sasaki for useful comments and discussions.
Chen and Romano are supported by the Taiwan NSC under Project No.\
NSC97-2112-M-002-026-MY3, by Taiwan's National Center for
Theoretical Sciences (NCTS). Chen is also supported by the US Department of Energy
under Contract No.\ DE-AC03-76SF00515.

\end{acknowledgments}

\end{document}